# Quantum well-based waveguide for semiconductor lasers

V Ya Aleshkin[1], A A Dubinov[1*], K E Kudryavtsev[1], A N Yablonskiy[1] and B N Zvonkov[2]

[1]Institute for Physics of Microstructures of the Russian Academy of Sciences,

Nizhny Novgorod, 603950, Russia

[2]Research Physical-Technical Institute of the Nizhny Novgorod State University,

Nizhny Novgorod, 603950, Russia

[*]sanya@ipm.sci-nnov.ru

## Abstract

In this work we study a possibility of waveguide fabrication on the basis of active quantum wells in semiconductor lasers. The efficiency of such a waveguide for an InP structure with $In_{0.53}Ga_{0.47}As$ quantum wells is demonstrated experimentally. An optically-pumped laser on this basis is realized.

## 1. Introduction

An intensive research is underway currently to upgrade the performance of semiconductor lasers through higher power, quantum efficiency and output radiation quality [1-3]. One very important component of a semiconductor laser, that is responsible for many of the laser characteristics, is a waveguide. A common practice in waveguide fabrication is to use either cladding layers with a refractive index smaller than in the waveguide core (like, for example, the InGaP or AlGaAs layers for GaAs based lasers), or a waveguide layer (InAlGaAsP) whose refractive index is higher than in the substrate (InP). In this case the substrate itself is a cladding layer.

As known, a symmetric waveguide (i.e., when the layer with a higher refractive index is sandwiched between unlimited layers with a lower refractive index) allows an existence of $TE_0$ and $TM_0$ modes, whatever its thickness [4]. Hence, an electromagnetic mode may be localized in the vicinity of a waveguide layer of thickness being several orders of magnitude smaller than the wavelength of this mode. For semiconductor lasers generating at wavelengths of about 1 μm even a ~10 nm thick layer may serve as such a waveguide. Note that similar thickness values are characteristic of quantum wells (QWs) that play the role of the active medium in lasers. Therefore, it is in principle possible to create a laser in which quantum wells will be both active and waveguide media. Such lasers do not require a conventional type of waveguide and are thus simpler of structure, which is very important technologically. However, in actual semiconductor lasers the waveguide can never be symmetric practically due to their fabrication technology that involves epitaxial growth of a laser structure and after-growth procedures, so that the grown quantum wells are located about a radiation wavelength away from the boundary with air (optical pumping) or metal (current pumping).

Besides, there is an optical confinement factor that determines the lasing threshold. Given very thin quantum wells and little difference between the refraction indices the optical confinement factor can be very small (weakly localized mode) and, hence, the lasing threshold too high. The objective of this work is to study a possibility of using quantum-size layers as a waveguide and to work out recommendations for construction of a laser with such waveguides.

## 2. Simple model

Consider a simple model describing the modes in a laser with a QW-based waveguide layer. In this case the radiation wavelength is largely in excess of the waveguide thickness, hence, the permittivity of the semiconductor structure with an electromagnetic wave propagating in it in the $x$ – direction can be described as follows:

$$\varepsilon(z) = \begin{cases} n_1^2 + \left(n_2^2 - n_1^2\right)d\delta(z), z < L \\ n_3^2, z \geq L \end{cases}, \quad n_2 > n_1 \geq n_3 \qquad (1),$$

where $n_1$, $n_2$, $n_3$, are the refraction indices of layers in the structure, $d$ and $L$ are the structure dimensions, $\delta(z)$ is the Dirac delta function.

The Maxwell equations for field $E_y(z,x)$ in the TE mode and field $H_y(z,x)$ in the TM mode can be written in the following form, respectively:

$$\frac{d^2 E_y(z,x)}{dx^2} + \frac{d^2 E_y(z,x)}{dz^2} + \varepsilon(z)\left(\frac{2\pi}{\lambda c}\right)^2 E_y(z,x) = 0 \tag{2},$$

$$\frac{d^2 H_y(z,x)}{dx^2} + \varepsilon(z)\frac{d}{dz}\left(\frac{1}{\varepsilon(z)}\frac{dH_y(z,x)}{dz}\right) + \varepsilon(z)\left(\frac{2\pi}{\lambda c}\right)^2 H_y(z,x) = 0 \tag{3},$$

where $\lambda$ is the mode wavelength, $c$ is the velocity of light in vacuum. At the boundary of the layers with different refractive indices the fields $E_y(z,x)$ and $H_y(z,x)$, $\dfrac{dE_y(z,x)}{dz}$ and $\dfrac{1}{\varepsilon(z)}\dfrac{dH_y(z,x)}{dz}$ are continuous. The boundary conditions for the waveguide modes are $E_y(z,x), H_y(z,x) \to 0$ at $z \to \pm\infty$. We will seek solution to equations (2) and (3) for field $E_y(z,x)$ in the TE mode and field $H_y(z,x)$ in the TM mode in the form:

$$E_y(z,x) = \exp(i\frac{2\pi}{\lambda}bx)\begin{cases} A_1 \exp(-i\frac{2\pi}{\lambda}N_1 z), z < 0, \\ B_1 \exp(i\frac{2\pi}{\lambda}N_1 z) + C_1 \exp(-i\frac{2\pi}{\lambda}N_1 z), 0 \leq z < L, \\ D_1 \exp(i\frac{2\pi}{\lambda}N_3 z), z \geq L \end{cases} \tag{4},$$

$$H_y(z,x) = \exp(i\frac{2\pi}{\lambda}bx)\begin{cases} A_2 \exp(-i\frac{2\pi}{\lambda}N_1 z), z < 0, \\ B_2 \exp(i\frac{2\pi}{\lambda}N_1 z) + C_2 \exp(-i\frac{2\pi}{\lambda}N_1 z), 0 \leq z < L, \\ D_2 \exp(i\frac{2\pi}{\lambda}N_3 z), z \geq L \end{cases} \tag{5},$$

respectively; here $A_1$, $B_1$, $C_1$, $D_1$, $A_2$, $B_2$, $C_2$, $D_2$ are the unknown constants, $b$ is the effective propagation index in the $x$ – direction, $N_1 = \sqrt{n_1^2 - b^2}$, $N_3 = \sqrt{n_3^2 - b^2}$. Then the dispersion equations for the TE and TM modes, where $b$ is unknown, have the form

$$\exp\left(i\frac{4\pi}{\lambda}N_1 L\right)(N_3 - N_1)\left(n_2^2 - n_1^2\right)d\frac{\pi}{\lambda} = (N_3 + N_1)\left(iN_1 + \frac{\pi}{\lambda}\left(n_2^2 - n_1^2\right)d\right) \tag{6},$$

$$\exp\left(i\frac{4\pi}{\lambda}N_1L\right)\left(\frac{N_3}{n_3^2}-\frac{N_1}{n_1^2}\right)\left(n_2^2-n_1^2\right)\frac{n_1^2}{n_2^2}d\frac{\pi}{\lambda}=\left(\frac{N_3}{n_3^2}+\frac{N_1}{n_1^2}\right)\left(iN_1+\frac{\pi}{\lambda}\left(n_2^2-n_1^2\right)\frac{n_1^2}{n_2^2}d\right) \quad (7),$$

respectively.

Let us now consider the behavior of the effective propagation index in this model for a particular structure of an $In_{0.53}Ga_{0.47}As$ layer of thickness $d$, grown on InP substrate and overgrown with an InP layer of thickness $L$. In this case for $\lambda = 1.55$ μm the refraction indices $n_1$ (InP) = 3.17 [5], $n_2$ ($In_{0.53}Ga_{0.47}As$) = 3.6 [6], $n_3$ (vacuum) = 1. Assume now that $b \to n_1$ and $N_1 \to 0$, which for the structure InP/$In_{0.53}Ga_{0.47}As$/InP means that $d \ll 75$ nm, $L \gg 0.2$ μm. Then, by expanding equations (6) and (7) in the small parameter $N_1$ we can derive the following equations for $b$ of the TE and TM modes, respectively:

$$\exp\left(-\frac{4\pi}{\lambda}\sqrt{b^2-n_1^2}\,L\right)+\frac{\lambda\sqrt{b^2-n_1^2}}{\pi d\left(n_2^2-n_1^2\right)}=1 \quad (8),$$

$$\exp\left(-\frac{4\pi}{\lambda}\sqrt{b^2-n_1^2}\,L\right)+\frac{\lambda\sqrt{b^2-n_1^2}}{\pi d\left(n_2^2-n_1^2\right)}\frac{n_2^2}{n_1^2}=1 \quad (9).$$

The solutions to these equations have the form:

$$b^{TE}=\left(n_1^2+\left[\frac{\pi}{\lambda}d\left(n_2^2-n_1^2\right)+\frac{\lambda}{4\pi L}W\left(-s\exp(-s)\right)\right]^2\right)^{\frac{1}{2}} \quad (10),$$

$$b^{TM}=\left(n_1^2+\left[\frac{\pi}{\lambda}\frac{n_1^2}{n_2^2}d\left(n_2^2-n_1^2\right)+\frac{\lambda}{4\pi L}W\left(-\frac{n_1^2}{n_2^2}s\exp\left(-\frac{n_1^2}{n_2^2}s\right)\right)\right]^2\right)^{\frac{1}{2}} \quad (11),$$

where $s=\left(\frac{2\pi}{\lambda}\right)^2 dL\left(n_2^2-n_1^2\right)$, $W(x)$ is the Lambert $W$-function [7].

Now, using the propagation indices found from formulae (10) and (11) in Eqs (4) and (5) we can estimate the optical confinement factor for the TE and TM modes in the waveguide structures of interest (assuming the waveguide and active layers dimensions equal):

$$\Gamma_{TE,TM} = \frac{\int_0^d |E_{y,z}(z,x)|^2 dz}{\int_{-\infty}^{\infty} |E_{y,z}(z,x)|^2 dz} \qquad (12),$$

where $E_z(z,x) = -\frac{bH_y(z,x)}{n(z)^2}$. The larger the optical confinement factor, the smaller gain is required to the laser generation threshold. In Fig. 1 the optical confinement factor is shown versus dimension $d$ of the waveguide layer for the TE and TM modes of InP/In$_{0.53}$Ga$_{0.47}$As/InP structure. For comparison we also show here the optical confinement factor in conventional waveguides. One can see that in the InP/In$_{0.53}$Ga$_{0.47}$As/InP structure with the active layer widths above 60 nm the value of the optical confinement factor tends to that of a conventional waveguide InP/In$_{0.73}$Ga$_{0.27}$As$_{0.58}$P$_{0.42}$/InP, which means that the fundamental mode in a regular waveguide is localized on a scale smaller than the waveguide width. This fact, in turn, implies that a laser based on the InP/In$_{0.53}$Ga$_{0.47}$As/InP structure with 6 quantum wells 10 nm wide each will have nearly the same factor of optical confinement as a laser with a wide waveguide layer In$_{0.73}$Ga$_{0.27}$As$_{0.58}$P$_{0.42}$, so, the latter layer is not necessary.

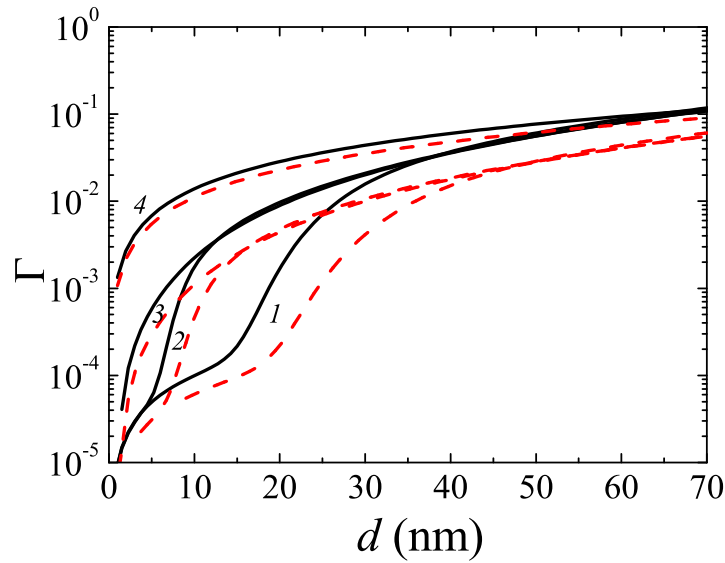

**Figure 1.** Optical confinement factor $\Gamma$ versus thickness $d$ of the active waveguide layer for TE (solid curve) and TM (dashed curve) modes and three values of $L$ (*1* - 1 μm, *2* - 3 μm, *3* – 20 μm) in the InP/In$_{0.53}$Ga$_{0.47}$As/InP structure (1.55 μm wavelength). Curves *4* are dependences of the optical con-

finement factor on the active layer thickness for TE (solid curve) and TM (dashed curve) modes for a conventional waveguide InP/In$_{0.73}$Ga$_{0.27}$As$_{0.58}$P$_{0.42}$/InP with a 1 μm thick InGaAsP layer.

## 3. The real structure model

In this section we consider a real model of semiconductor laser with the active medium ($d = pd_{QW}$) of several ($p$ = 1 - 6) identical quantum wells of thickness $d_{QW}$ = 10 nm, separated by 20 nm wide identical barriers. For a current-pumped laser it is also necessary to take into account the contact of the structure with metal and the doping of substrate (2×10$^{18}$ cm$^{-3}$ donor concentration) and contact layer (2×10$^{18}$ cm$^{-3}$ acceptor concentration). Note that the latter produces a waveguide effect for the TE and TM modes, as the refractive index of semiconductor is somewhat reduced through the doping process. However, such waveguides feature considerable losses due to the field penetration and free-carriers absorption in the doped regions of semiconductor. Waveguides of this kind were used in the first diode lasers with the *p-n* homojunction [3]. The waveguiding effect of doping gets stronger with an increasingly smaller value of such quantities as the difference in the refractive indices of quantum wells and the semiconductor matrix (in which QWs are located), the thickness and number of QWs. A waveguide effect for the TM mode may also occur through excitation of a surface plasma wave in the laser at the semiconductor-metal interface [8]. Yet, since this wave has big losses in the near-IR range, we did not consider it in this work.

The electric field distribution, propagation indices and optical confinement factors were found numerically from Eqs (2), (3), (12) by the transfer-matrix method for both structures in either situation. The doping effect on the semiconductor refractive index was taken into account by the following formula [9]:

$$n_c = \sqrt{n^2 - \frac{4\pi e^2 N_c}{m^* \omega^2}} \quad (13),$$

where $n_c$ and $n$ are the refractive indices of doped and undoped semiconductor, respectively, $e$ is the electron charge, $N_c$ the concentration of charge carriers, $\omega$ is the radiation frequency, $m^*$ is the effective mass of a charge carrier, taken from the *Data Handbook* [10].

Dependence of the optical confinement factor on the active medium size (number of quantum wells) for the TE and TM modes and two values of $L$ in the InP/In$_{0.53}$Ga$_{0.47}$As/InP structure is shown in Fig. 2. One can see that the presence of barriers between quantum wells in the active medium of laser based on this structure does not practically affect the value of the optical confinement factor. However, for an exact solution there is a threshold for TE and TM modes in this structure ($L = 1$ μm), that for the TM mode (3 QWs) is higher than for the TE mode (2 QWs). The contact of structure with metal and the doping of substrate and contact layer affect the TE mode only given few quantum wells (fewer than 1-2) and small $L$. The TM mode, on the contrary, was found to be largely influenced by the contact of structure with metal at small values of $L$. This led us to a conclusion that the TM mode in a structure with $L = 1$ μm exists only if the number of quantum wells in the active medium is larger than 5.

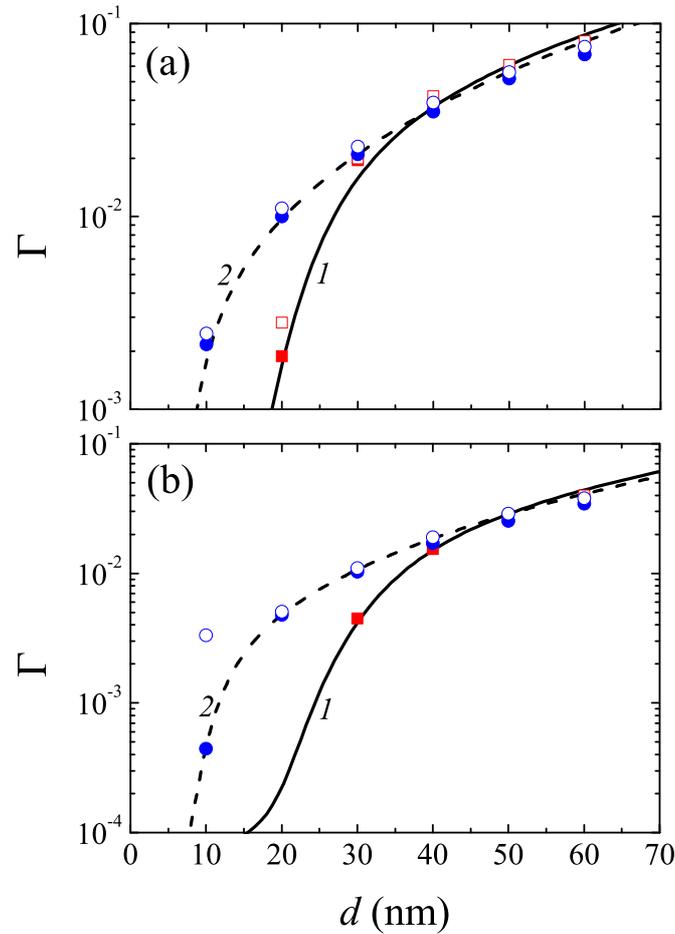

**Figure 2.** Optical confinement factor $\Gamma$ versus thickness $d$ of the active waveguide layer for TE (a) and TM (b) modes and two values of $L$ (*1* - 1 μm, *2* - 3 μm) in the InP/In$_{0.53}$Ga$_{0.47}$As/InP structure at the 1.55 μm wavelength, calculated in the simple model (solid and dashed curves) by formulas (10) and (11). Shaded squares and circles correspond to the optical confinement factors for the TE and TM modes, found by rigorous calculation for an active medium of $p$ ($p$ = 1 - 6) quantum wells of thickness $d_{QW}$ = 10 nm ($d = pd_{QW}$), that are separated by 20 nm thick barriers. Hollow squares and circles indicate the optical confinement factors for the TE and TM modes for the same quantum wells with account of doping and $n_3$ (gold) = 0.5 + 10$i$ [11].

## 4. Experiment

To experimentally prove the proposed theory we investigated an InP/In$_{0.53}$Ga$_{0.47}$As structure with three quantum wells grown by the MOCVD method in a horizontal reactor at atmospheric pressure. The In$_{0.53}$Ga$_{0.47}$As quantum wells 10 nm thick each, separated by 100 nm wide InP barriers were grown on the InP substrate. They were then overgrown with an InP layer of 1300 nm thickness. The thinned structures were split into 1 mm wide strips, the (110) facet cleavages were used as mirrors.

The threshold of stimulated emission from the structure was reached (Fig. 3) at ~ 260 W/cm$^2$ power density by pumping with a cw frequency-doubling Nd:YAG laser (532 nm wavelength) at liquid nitrogen temperature (77 K), which provides evidence of the waveguide effect of quantum wells. The generation threshold at room temperature (293 K) was much higher (~ 5 kW/cm$^2$ by pumping with an MOPO-SL parametric generator of light ('Spectra-Physics') having 530 nm wavelength, ~ 10 ns pulse duration and 10 Hz pulse repetition rate). A diode array (0.62-2.2 μm operating range) was used as the radiation detector. Such a big difference in the generation threshold values at 77 K and 293 K depends, according to [12], on a considerable increase in the frequency of Auger recombination with a growing temperature for the In$_{0.53}$Ga$_{0.47}$As quantum wells that are lattice-matched with InP. It is also shown in [12] that by using strained quantum wells it is possible to largely reduce the generation threshold at room temperature.

Note that the laser radiation could be observed only from the cleaved facet of the structure. It means that there is practically no radiation scattering in the QW-based waveguide and, so, its waveguiding properties are good.

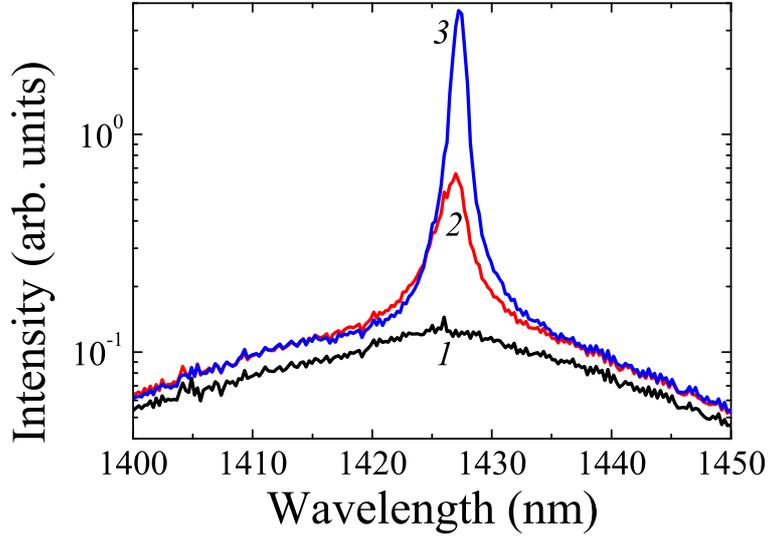

**Figure 3.** Emission spectrum from an InP/In$_{0.53}$Ga$_{0.47}$As structure with 3 quantum wells for three values of power density by pumping with a cw frequency-doubling Nd:YAG laser (532 nm wavelength) at T = 77 K: *1* – 250 W/cm$^2$, *2* – 260 W/cm$^2$, *3* – 315 W/cm$^2$.

### 5. Conclusion

In this paper we have considered a model problem of a semiconductor laser with a waveguiding quantum-size layer, and a real model taking into account the number of barrier-separated quantum wells and the doping of substrate and semiconductor-metal contact layer. It was shown that TE and TM modes may be localized near a quantum well. The mode localization depends for its magnitude on a difference in the refractive indices of quantum well and semiconductor matrix, on a quantum well thickness and on a distance between the quantum well and the semiconductor/vacuum or semiconductor/metal interface: the larger these quantities, the larger the mode localization. A possibility of effectively using such waveguides for laser structures characterized by a large number of quantum wells (for, example, InP/In$_{0.53}$Ga$_{0.47}$As/InP) and a considerable difference in the refractive indices of the quantum well and semiconductor matrix materials has been demonstrated. An optically-pumped laser of this kind has been practically realized.

## Acknowledgment

This work was supported by the RFBR (12-02-90024-Bel, 11-02-97049-Povolzhje), Grants of the President of the Russian Federation for support of young Russian scientists (MK-678.2012.2) and leading scientific schools (HIII-4756.2012.2), the Ministry of Education and Science (project 2012-1.2.1-12-000-2013-7607).